\documentclass[aip,reprint]{revtex4-1}

\usepackage{graphicx}
\usepackage{mathtools}
\usepackage{bm}
\usepackage{amsmath}
\usepackage{amssymb}

\draft 

\begin{document}

\title{Static force characterization with Fano anti-resonance in levitated optomechanics}

\author{Chris Timberlake}
\email{ct10g12@soton.ac.uk}
\affiliation{Department of Physics and Astronomy, University of Southampton, Southampton
SO17 1BJ, UK}

\author{Marko Toro\v{s}}
\email{m.toros@soton.ac.uk}
\affiliation{Department of Physics and Astronomy, University of Southampton, Southampton
SO17 1BJ, UK}

\author{David Hempston}

\affiliation{Department of Physics and Astronomy, University of Southampton, Southampton
SO17 1BJ, UK}

\author{George Winstone}

\affiliation{Department of Physics and Astronomy, University of Southampton, Southampton
SO17 1BJ, UK}

\affiliation{School for Materials Science, Japan Advanced Institute of Science and Technology, Nomi, Ishikawa 923-1211, Japan}

\author{Muddassar Rashid}

\affiliation{Department of Physics and Astronomy, University of Southampton, Southampton
SO17 1BJ, UK}

\author{Hendrik Ulbricht}
\email{h.ulbricht@soton.ac.uk}
\affiliation{Department of Physics and Astronomy, University of Southampton, Southampton
SO17 1BJ, UK}

\date{\today}
\begin{abstract}
We demonstrate a classical analogy to the Fano anti-resonance in levitated optomechanics by applying a DC electric field. Specifically, we experimentally tune the Fano parameter by applying a DC voltage from 0~kV to 10~kV on a nearby charged needle tip. We find consistent results across negative and positive needle voltages, with the Fano line-shape feature able to exist at both higher and lower frequencies than the fundamental oscillator frequency. We can use the Fano parameter to characterize our system to be sensitive to static interactions which are ever-present. Currently, we can distinguish a static Coulomb force of $2.7 \pm 0.5 \times 10^{-15}$~N with the Fano parameter, which is measured with one second of integration time. Furthermore, we are able to extract the charge to mass ratio of the trapped nanoparticle.
\end{abstract}
\maketitle

Resonance is of significant importance in a wide variety of fields within Physics, with the phenomena found in both classical and quantum systems. In 1961, Fano discovered that, in optics, an asymmetric line-shape arises by the interference between a discrete localized state and a continuum of states~\cite{Fano1961,Fano1935}.

Fano interference has been demonstrated in numerous quantum mechanical systems including semiconductor nanomaterials~\cite{Holfeld1998,Fan2014}, quantum wells~\cite{Faist1997,Nikonov1999} and quantum dots~\cite{Kobayashi2002,Johnson2004,Sasaki2009}, superconductors~\cite{Limonov1998,Hadjiev1998}, dielectric~\cite{Chong2014} and gold nanoparticles~\cite{Stockman2010}, photonic crystals ~\cite{Miroshnichenko2010,Soboleva2012,Yang2012,Rybin2010,Zhou2014,Markos2015,Limonov2017a}, electromagnetically induced transparency (EIT) in interactions between three-level atomic systems and two laser fields~\cite{Marangos1998,Fleischhauer2005}, and many other examples.

Although the Fano anti-resonance phenomena has been widely acknowledged as an effect in quantum systems, it is a general wave phenomena, meaning it can manifest itself in numerous classical systems also. Studies on the classical interpretation of the Fano effect have been undertaken~\cite{Joe2006,Satpathy2012}, as well as theoretical comparison of the quantum and classical Fano parameter~\cite{Iizawa2018}. Experimental evidence of Fano anti-resonance has been shown for classical nanomechanical oscillators~\cite{Tribelsky2008,Liu2009,Hao2008,Verellen2009,Stassi2017}, whispering-gallery microresonators~\cite{Li2011} and in prism-coupled square micro-pillars~\cite{Lee2004}.

Levitated nanoparticles have recently emerged as very promising candidates for measuring extremely small forces. This is typically done by measuring the resonant response to a perturbation on its motion~\cite{Ranjit2015, Ranjit2016, Hempston2017}. Levitated systems have also been used to study interactions with nearby dielectric surfaces~\cite{Winstone2018, Diehl2018}, and proposals have been devised to measure short range interactions like the Casimir effect~\cite{Geraci2010}. Recently Hebestreit et al. proposed and demonstrated detection of static forces using free falling nanoparticles, with a sensitivity of 10~aN reported~\cite{Hebestreit2018}. 

In this letter we experimentally demonstrate Fano anti-resonance in levitated optomechanics. The characteristic Fano anti-resonance is induced with a static Coulomb interaction by charging a stainless steel needle close to a charged nanoparticle in a gradient force optical trap. We show we can tune the Fano parameter by varying the voltage applied to the needle tip, and use the asymmetry in the line-shape to characterize a method of static force detection. The advantage of this method is precise force sensing irrespective of the resonant frequency. Consistent Fano parameter results were found for positive and negative applied voltages. In addition we are able to extract the charge to mass ratio of the trapped nanoparticle. We also give a phenomenological model to describe the Fano line-shape and extract a characteristic rate.

\begin{figure}[t!]
\centering 
\includegraphics[width=1\linewidth]{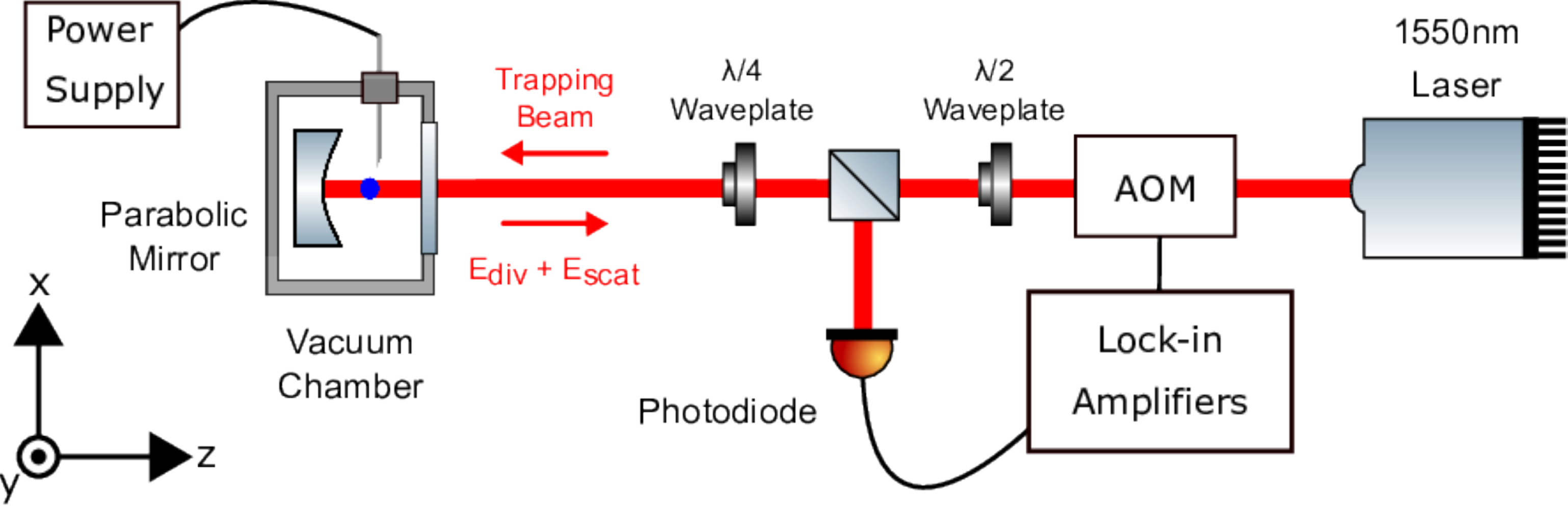}
\caption{Experimental setup. A silica nanoparticle is trapped, and detected, with a 1550~nm laser at the focus of a paraboloidal mirror. An AOM is used to modulate the laser power and cool the motion of the nanoparticle. A nearby stainless steel needle is connected to a power supply to apply a voltage which is used to manipulate the motion of a charged nanoparticle in the trap.}
\label{fig:Setup} 
\end{figure}

We consider a polarizable nanoparticle trapped in an optical trap
which has been described in~\cite{torovs2018detection}. It is convenient
to define the center of the optical trap as the origin of the coordinate
system: the $x$ axis is the vertical direction pointing away from
the ground, the $z$ axis is oriented in the direction of the beam
propagation away from the mirror, and the $y$ axis is the remaining
horizontal axis (see Fig.~\ref{fig:Setup}). To stabilize the motion
at low pressure $p$ the nanoparticle is cooled using parametric feedback
cooling~\cite{gieseler2012subkelvin,Vovrosh2017,setter2018real}.
The strong cooling confines the motion of the nanoparticle to small
oscillations, $\delta\bm{r}=(\delta x,\delta y,\delta z)^{\top}$,
around an equilibrium position, $\bm{r}_{\text{eq.}}=(x_{0},y_{0},z_{0})^{\top}$,
which effectively decouples the translational motions. As we will
discuss below, the scattering force $\mathbf{F}_{s}=(0,0,F_{\text{scatt}})$
displaces the equilibrium point $\bm{r}_{\text{eq.}}$ away from the
mirror, i.e.~$x_{0},\,y_{0}\ll z_{0}$, which makes the $z$-dynamics
independent of the $x_{0}$ and $y_{0}$ values. In the following
we limit the discussion to the $z$-motion as it is experimentally
the strongest signal. 

The potential generated by the optical field can be modeled using
the following potential:
\begin{equation}
U_{\text{opt}}(z) =\frac{m}{2}\omega_{0}^{2}z^{2}-\eta z^{4}, \label{eq:opt}
\end{equation}
where $m$ is the mass of the nanoparticle, $\omega_{0}^{2}=\frac{2P\chi}{c\sigma_{L}\rho z_{R}^{2}}$
\cite{rashid2017wigner}, $P$ is the laser power, $\sigma_{L}=\pi w_{0}^{2}$
is the effective laser beam cross section area, $w_{0}$ is the mean
beam waist radius, $z_{R}$ is the Rayleigh length, $\rho$ is the
particle density, $\chi$ is an electric susceptibility of the particle, $c$ is the speed of light, and $\eta$ quantifies the dominant
non-linearity of the trap~\cite{Gieseler2013}. 

The scattering force has only one non-zero component $F_{\text{scatt}}(z)$
which is given in~\cite{Rashid2018}. We integrate it,
i.e.~$\int dz'F_{\text{scatt}}(z')$, which gives the following effective
potential: 
\begin{equation}
U_{\text{scatt}}(z)=-\frac{32\pi^{3}\hbar\text{\ensuremath{\Gamma_{s}}}w_{0}^{2}\tan^{-1}\left(\frac{\lambda z}{\pi w_{0}^{2}}\right)}{3\lambda^{2}},\label{eq:scatt}
\end{equation}
where $\Gamma_{s}=\frac{\sigma_{R}}{\sigma_{L}}\frac{P}{\hbar\omega_{L}}$
is the scattering rate, $\sigma_{R}=\frac{\pi^{2}V_{0}^{2}}{\lambda^{4}}$
is an effective cross-section area, $\omega_{L}=\frac{2\pi c}{\lambda}$,
and $\lambda$ is the wavelength of light. 

Consider now also a nearby charged needle which generates a Coulomb
potential $\propto1/|\bm{R}-\bm{r}|,$ where $\mathbf{R}=(\frac{1}{\sqrt{2}}R,0,\frac{1}{\sqrt{2}}R)$
is the position of the needle tip. For the experimental situation
described in this paper only the linear contribution is relevant,
specifically, expanding the Coulomb potential to and including order
$\mathcal{O}(z)$ one readily finds
\begin{equation}
U_{\text{el}}(z)=\frac{qQz}{4\pi\epsilon_{0}\sqrt{2}R^{2}},\label{eq:el}
\end{equation}
where $Q$ is the charge on the needle tip, $q$ is the charge on
the nanoparticle, and $\epsilon_{0}$ is the permittivity of free
space.

In addition, we apply sinusoidal modulations of the laser power $P$
to cool the center-of-mass motion (c.m.) of the nanonparticle, namely, parametric
feedback cooling~\cite{Vovrosh2017, gieseler2012subkelvin, setter2018real}. In a nutshell, one can cool a translational degree
of freedom by tracking its phase (in classical phase space) and applying
a modulation at twice its harmonic frequency. Specifically, the feedback
term can be obtained by making the formal replacement $P\rightarrow P(1+\beta zp_{z})$
in the equations of motion, where $p_{z}$ denotes the conjugate momentum,
and $\beta$ is the strength of the feedback which has units of the
inverse of an action, i.e. $\text{kg}^{-1}\text{m}^{-2}\text{s}$.
This procedure generates the following feedback force term in the
dynamics~\cite{setter2018real}:
\begin{equation}
f_{\text{fb}}=\beta\partial_{z}(U_{\text{opt}}+U_{\text{scatt}})zp_{z}.\label{eq:feedback}
\end{equation}
Taking into account the feedback term one obtains the following dynamics:
\begin{ruledtabular}
\begin{alignat}{2}
\dot{z} & =\frac{z}{m},\label{eq:rdot}\\
\dot{p}_{z} & =-{\partial_{z}}U_{\text{eff}}-f_{\text{fb}}-2\gamma_{\text{coll}}p_{z}+m\xi,\label{eq:pdot}
\end{alignat}
where the total effective potential is given by $U_{\text{eff}}=U_{\text{opt}}+U_{\text{scatt}}+U_{\text{el}}$.
We have also included a damping term with coupling $\gamma_{\text{coll}}$,
which is the gas collisions rate, and a noise term $\xi$.

We now combine Eqs.~\eqref{eq:rdot} and~\eqref{eq:pdot}, using
$\dot{p}_{z}=m\dot{z}$, and Taylor expand around the minimum position
$z_{0}$, up to and including order $\mathcal{O}(\delta z^{2})$.
Specifically we obtain the term $m\omega_{\text{m}}^{2}\delta z^{2}/2$,
where the harmonic frequency is given by
\begin{equation}
\omega_{\text{m}}=\sqrt{\text{\ensuremath{\omega_{0}^{2}-\eta\frac{12z_{0}^{2}}{m}+}}\frac{64\pi^{4}\lambda w_{0}^{4}\hbar z_{0}}{3m\left(\pi^{2}w_{0}^{4}+\lambda^{2}z_{0}^{2}\right)^{2}}\Gamma_{s}}.\label{eq:omega_m}
\end{equation}
Eq.~\eqref{eq:omega_m} describes how the measured frequency $\omega_{\text{m}}$
is related to the dipole-trap harmonic frequency $\omega_{0}$, the
trap non-linearity term $\propto\eta$, and the scattering force term
$\propto\Gamma_s$. Here we neglect other smaller effects that could change
the particle's frequency.

We are left to discuss the noise term $\xi$. Suppose the noise is
invariant under time-translations, has zero mean, and is fully quantified
by the two point correlation function $f(\tau)=\mathbb{E}[\xi(t+\tau)\xi(t)]$,
where $\mathbb{E}[\,\cdot\,]$ denotes the average over different
noise realizations. Exploiting the Wiener\textendash Khinchin theorem~\cite{Chatfield1989}
one then readily finds the power spectral density of the $z$-degree
of freedom: $S_{zz}(\omega)\propto\tilde{f}(\omega)/(\omega^{2}\Gamma^{2}+(\omega^{2}-\omega_{\text{m}}^{2})^{2})$,
where $\tilde{f}$ denotes the Fourier transform of $f(\tau)$, and
$\Gamma$ is an effective damping rate comprising $\gamma_{\text{coll}}$
as well as an additional damping contribution due to feedback term.
Furthermore, assume that $\tilde{f}$ has two distinct noise sources:
one related to gas collisions $\propto\gamma_{\text{coll}}$, and
one related to the Coulomb force $\propto qQ$. The term related to
gas collisions, which is commonly present in optomechanical systems,
leads to the usual Lorentzian power spectral density: 
\end{ruledtabular}
\begin{equation}
S_{\text{coll}}(\omega)\propto\frac{\gamma_{\text{coll}}}{\omega^{2}\Gamma^{2}+(\omega^{2}-\omega_{\text{m}}^{2})^{2}}.\label{eq:Scoll}
\end{equation}
Here, we also find a noise source $\propto qQ$. Here we limit the analysis to
a phenomenological description of the effect, leaving a proper derivation
for future work. We make the following ansatz for the power spectral
density associated to the noise perturbed Coulomb force
\begin{equation}
S_{\text{el}}(\omega)\propto\frac{(\mathcal{-\mathfrak{f}}\gamma_{\text{el}}^{2}+(\omega^{2}-\omega_{\text{m}}^{2}))^{2}}{\omega^{2}\Gamma^{2}+(\omega^{2}-\omega_{\text{m}}^{2})^{2}},\label{eq:Sel}
\end{equation}
where $\mathfrak{f}=\pm e_{0}^{2}/(qQ)$ is a number, $e_{0}$ denotes
the unit charge of one electron, and $\gamma_{\text{el}}$ is a characteristic
rate. This ansatz has been inspired by the asymmetric Fano line-shape~\cite{Fano1961} and we will refer to $\mathfrak{f}$ as the Fano parameter.
We speculate that $S_{\text{el}}(\omega)$ could originate from a
non-Markovian noise, self-induced by the motion of the nanoparticle,
i.e. the motion of the nanoparticle could be perturbing the charges
on the needle tip and on the particle, which would perturb back the
motion of the nanoparticle on a time scale $\propto\mathfrak{f}^{-1/2}\gamma_{\text{el}}^{-1}$. Another possibility is the noise could originate from vibrations in the silica nanoparticle \cite{Heiman1979}.

Adding all the contributions from Eqs.~\eqref{eq:Scoll} and~\eqref{eq:Sel}
we finally arrive at
\begin{equation}
S_{zz}(\omega) = A+BS_{\text{coll}}(\omega)+CS_{\text{el}}(\omega),\label{eq:Fano_fit}
\end{equation}

\noindent where $A$, $B$, and $C$ are free parameters, i.e. fitting
constants which account for the finite noise floor and normalizations.
The order of magnitude of $A$ and $B$ can be first fixed when the
charge on the needle is absent. This leaves two free
parameters, $C$ and $\gamma_{\text{el}}$. 

The experimental setup consists of an optical gradient force trap, which is generated by tightly focusing a 1550~nm laser with a high numerical aperture (N.A.) paraboloidal mirror~\cite{Vovrosh2017,Rashid2016a,Hempston2017}. Silica nanoparticles of radius $\sim 75$~nm (density $\sim1800\ \text{kg}/\text{m}^{3}$) are prepared in a water based solution before being dispersed into a vacuum chamber, where a single particle is trapped at the focus. The position of the trapped nanoparticle is measured by detecting the interference generated between the Rayleigh scattered light from the particle, $E_\text{scat}$, and the divergent reference field, $E_\text{div}$, at a single photodiode (as shown in Fig.~\ref{fig:Setup}). The detected signal consists of information of three distinct translational frequency modes, which we refer to as $x$, $y$ and $z$. These three modes are tracked using lock-in amplifiers, which feeds the information to an acousto-optic modulator (AOM) which applies a modulation to the laser intensity to parametrically cool the c.m.~motion of the nanoparticle. A stainless steel needle, which is connected to a high voltage power supply, is placed close to the trapping region allowing us to influence the motion of a charged nanoparticle via the Coulomb interaction. The fine tip of the needle results in the charge being concentrated on the needle tip, meaning the charged needle can be considered to be a point charge~\cite{Hempston2017}. 

\begin{figure}[t!]
\centering 
\includegraphics[width=1\linewidth]{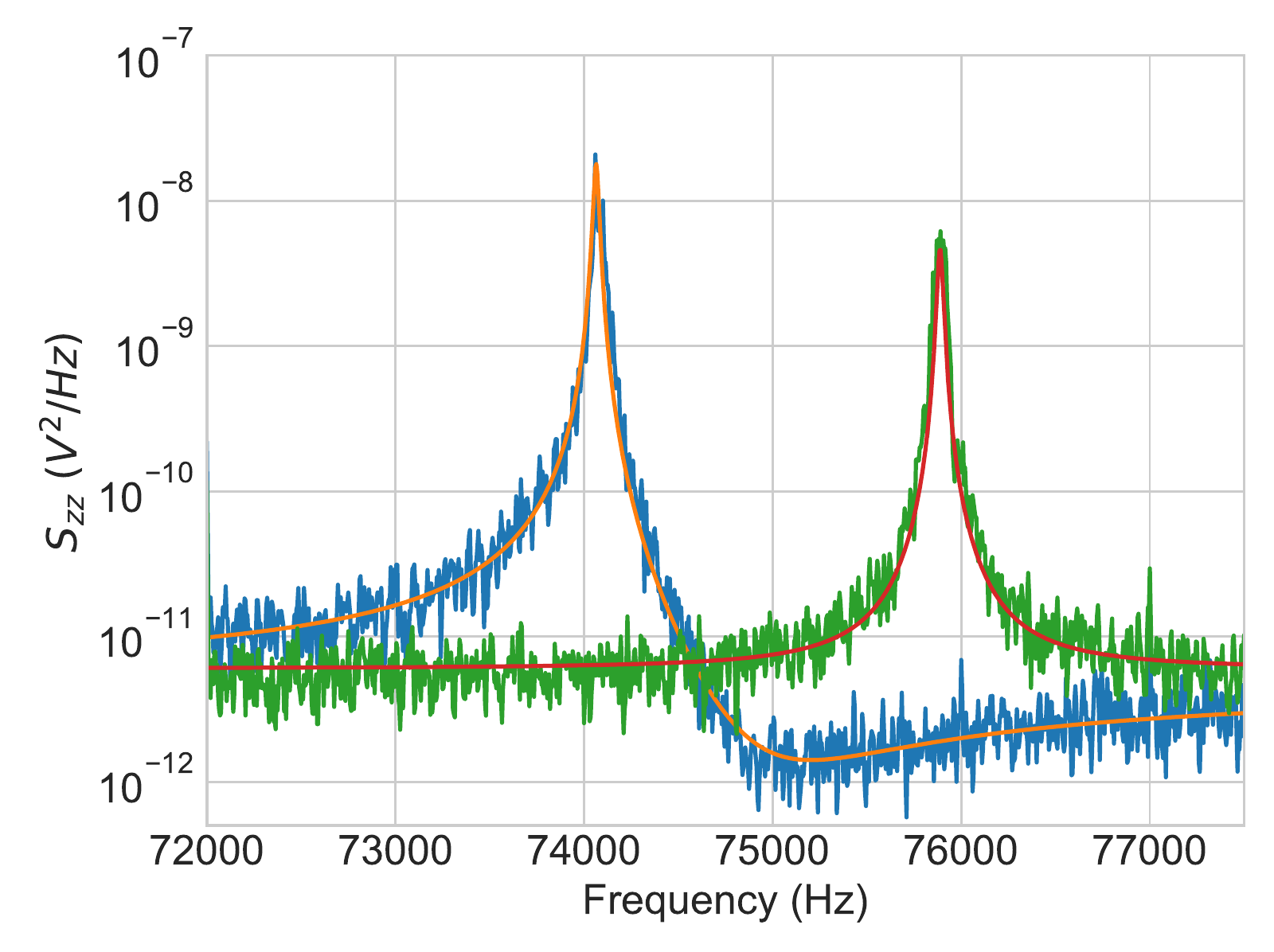} \caption{The power spectral density of a levitated nanoparticle with 0~kV applied to the needle (green) and 10~kV (blue). The 0~kV peak is fitted with a standard Lorentzian distribution from Eq.~\eqref{eq:Scoll} while the 10~kV peak is fitted with the asymmetric Fano line-shape from Eq.~\eqref{eq:Fano_fit}.}
\label{fig:10kV_0kV} 
\end{figure}

The voltage applied to the needle is swept from 0~kV to 10~kV, with one second of data taken for each voltage interval, and the effect this has on the shape of the $z$ frequency spectrum is shown in Fig.~\ref{fig:10kV_0kV}. Due to the anti-resonance suppressing the mechanical noise, there is also a reduction in the noise floor compared to the standard Lorentzian; for an applied voltage of 10~kV this reduction is approximately a factor of five for a small bandwidth around 1~kHz from the $z$ frequency.

\begin{figure}[t!]
\centering 
\includegraphics[width=1\linewidth]{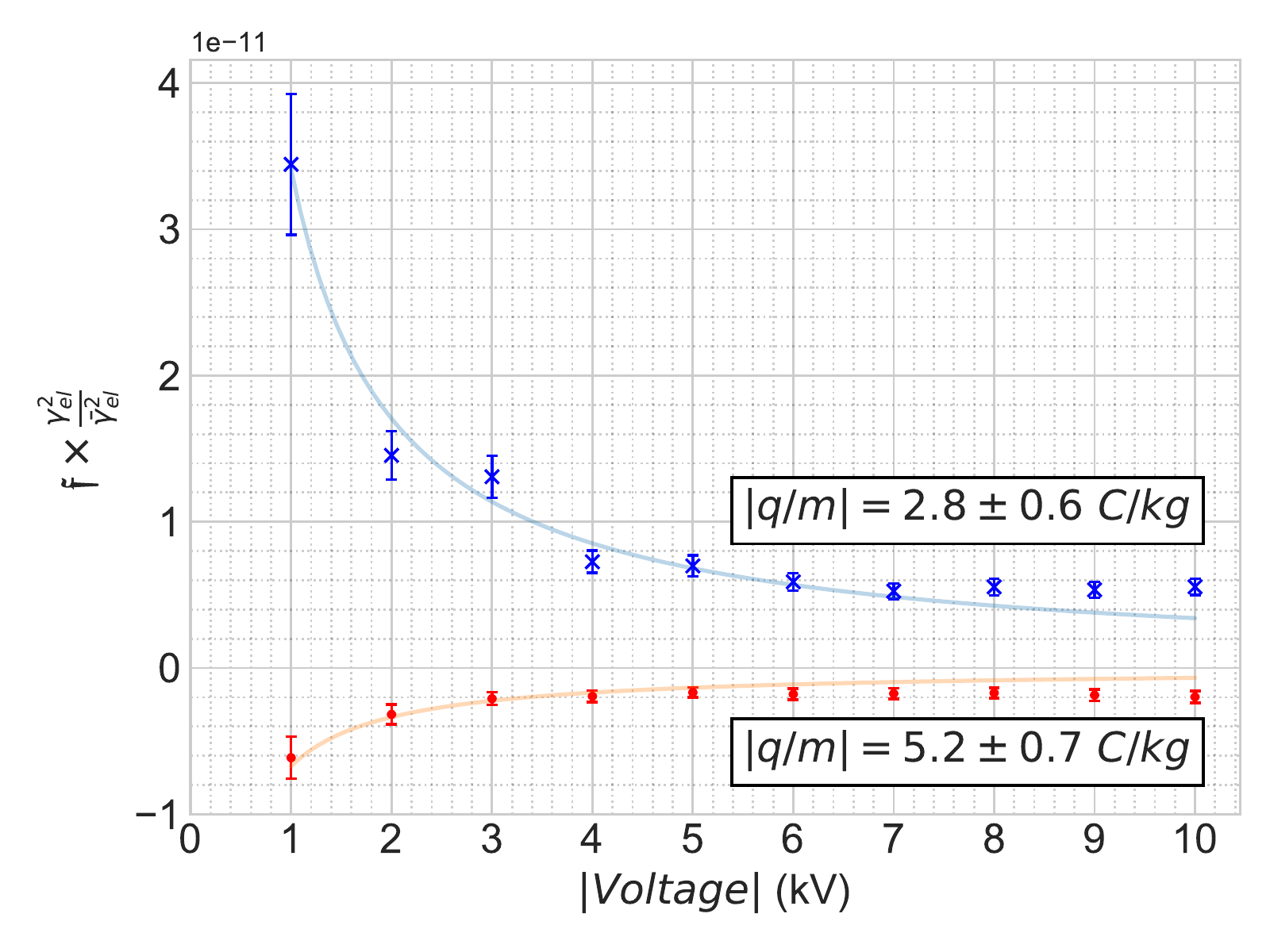}
\caption{The Fano parameter $\mathfrak{f}$ for two particles as a function of applied voltage, extracted using Eq.~\eqref{eq:Fano_fit}. The data was taken at a vacuum pressure of $8\times10^{-5}$~mbar and $3\times10^{-5}$~mbar for the blue and red data, respectively. The Fano parameter $\mathfrak{f}$ scales with the inverse of voltage, which is in good agreement with the theory.}
\label{fig:Fano_parameter} 
\end{figure}

We have fitted Eq.~\eqref{eq:Fano_fit} to data for two trapped particles in Fig.~\ref{fig:Fano_parameter}, exploring the regime of both the positive and the negative Fano parameter, for which we find an average characteristic rate, $\bar{\gamma}_{\text{el}}$, of $3.2 \pm 0.3$~GHz (positive Fano parameter) and $3.7 \pm 0.8$~GHz (negative Fano parameter), respectively. We find excellent agreement at low voltages ($<$ 5~kV), with slight deviation at higher voltages ($>$ 5~kV) which might be due to additional noise sources such as the ripple voltage in the high voltage power supply.

\begin{figure}[t!]
\centering \includegraphics[width=1\linewidth]{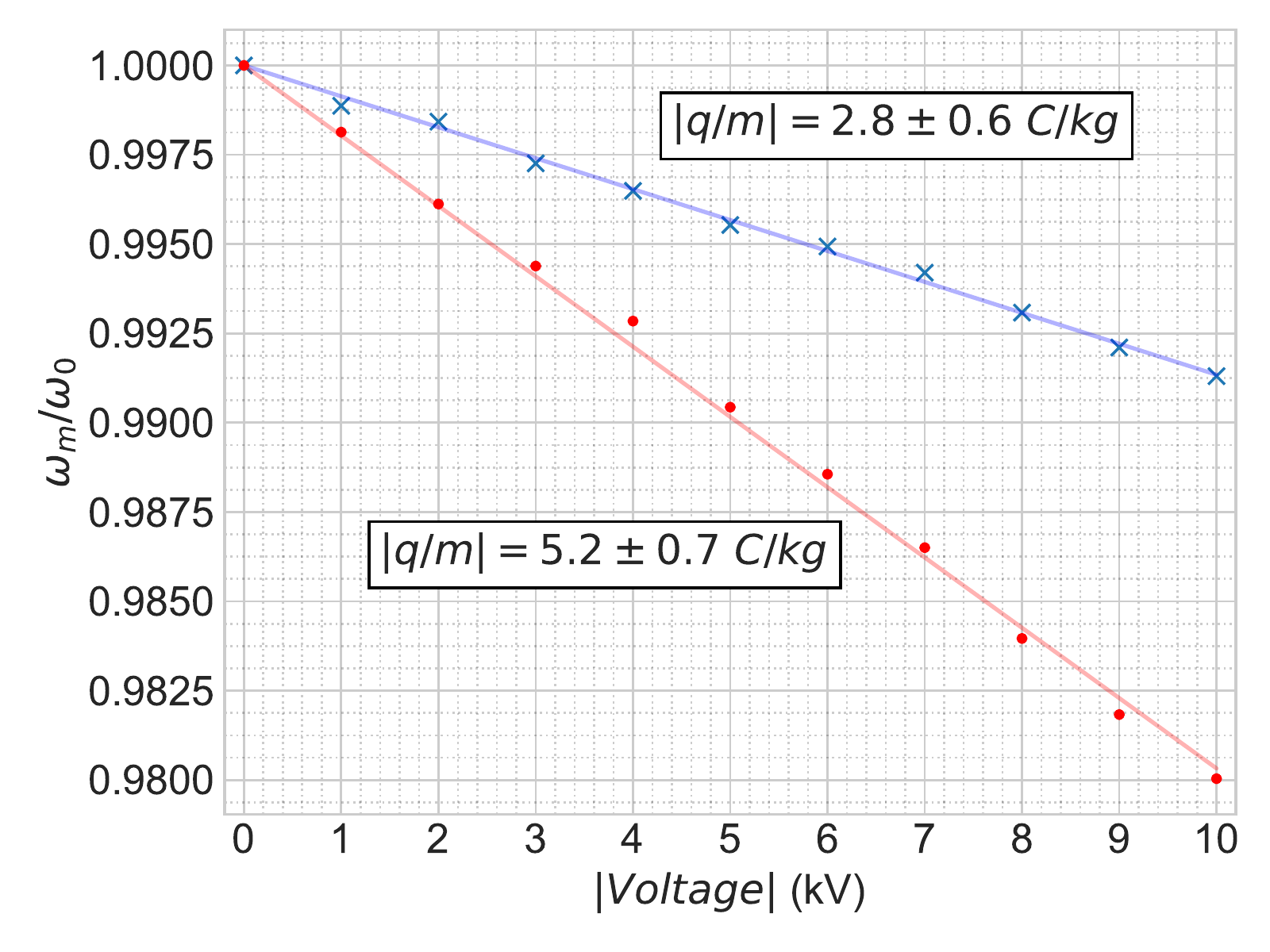} \caption{Frequency shift of the $z$ motion as a function of applied voltage to the needle, fitted using Eq.~\eqref{eq:omega_m}. The applied voltage displaces the equilibrium position of the particle's motion, which changes the total potential $U_{\text{eff}}$ the particle experiences. This results in a frequency shift which is linear with increasing voltage, to first approximation. Here two particles are plotted to show the linear response to the applied voltage. It can be seen that a larger charge to mass ratio results in a larger relative frequency shift.}
\label{fig:Volts_Freq} 
\end{figure}

The application of an electric field also results in a shift in the average position $z_0$.  Specifically, in the experimentally considered regime we find an approximate expression for the minimum position $z_0\approx-\frac{1}{m\omega_0^2} \left(\frac{qQ}{ 4 \pi \epsilon_0 \sqrt{2} R^2}-\frac{32 \pi ^2 \Gamma_s \hbar }{3 \lambda }\right)$. Furthermore, as the laser intensity is different in the new equilibrium position $z_0$ there is also a corresponding shift in oscillation frequency  $\omega_\text{m}$. This frequency shift is generally highly dependent on both the scattering force and the non-linearities in the gradient force potential~\cite{Gieseler2013}, as can be seen in Eq.~\eqref{eq:omega_m}. However, limiting again the discussion to the regime where non-linearites can be neglected we find a simplified expression for the frequency shift $
\omega_\text{m}\approx \omega_0 \left(1+  \frac{32 \Gamma_S \lambda   \hbar }{3  w_0^4   }\frac{z_0}{m \omega_0^2}\right)$, where $\omega_0$ is given below Eq.~\eqref{eq:opt}. Combing the formulas in this paragraph, as well as exploiting the formula for the scattering rate $\Gamma_S\propto m^2$ below Eq.~\eqref{eq:scatt}, we obtain the following simplified expression for the mechanical frequency 
\begin{equation}
\omega_\text{m}=\omega_0+\mathcal{B} m^2+\mathcal{C} q Q, \label{fiteq}
\end{equation}
where $\mathcal{B}=(\frac{16\pi}{3}\frac{P}{\omega_0^{3/2}w_0^4 c \lambda^3 \rho^2})^2$ and $\mathcal{C}=\frac{2\sqrt{2} P}{3\pi \epsilon_0 R^2 \lambda^2 w_0^6 \omega_0^3 \rho^2 c}$ depend only on the trapping laser field, the optics,  and the intrinsic particle properties, i.e. density $\rho$ and the electric susceptibility $\chi$; all these parameters can be estimated independently. Furthermore, we can also estimate the charge $Q$ on the needle tip from the applied voltage $V$ using COMSOL Multiphysics; we have tested the reliability of the numerical simulation by comparing with analytical methods used in previous works~\cite{Hempston2017} and, since we can set the desired voltage to high precision, we can in first approximation neglect the error on the estimated value of $Q$. By fitting Eq.~\eqref{fiteq} we can then extract the values of the mass $m$ and charge $q$. The former value, $m$, is in good agreement with the mass we  obtain by fitting to the frequency spectrum of the particle~\cite{Vovrosh2017, gieseler2012subkelvin}. From the values of $q$ and $m$ we can then estimate the charge to mass ratio $q/m$. The frequency shift due to the varying voltage for two different particles can be seen in Fig. \ref{fig:Volts_Freq}.

Static perturbations which cannot be \char`\``{}switched off\char`\"{} are typically hard to characterize because the oscillator experiences this static effect at all times, whereas the Fano parameter induced by such a perturbation will be present to allow characterization without \char`\``{}switching off\char`\"{}, or varying, such perturbation. For example, forces such as those induced by radiation pressure or the Earth's gravitational attraction could be probed using this technique. Here we use the Coulomb interaction to characterize how sensitive our system is to such static fields.  

From the experimental data in Fig.~\ref{fig:Volts_Freq} we can also extract the magnitude of the Coulomb force by fitting again Eq.~\eqref{fiteq} as discussed above: here we interested only in the extracted value of the charge product, $q Q$. We can then obtain the Coulomb force given by $F=\frac{1}{4\pi\epsilon_0 }\frac{q Q}{R^2}$, as we can measure $R$ independently. In the following we discuss results for a particle of radius $71 \pm 11$~nm, which has a charge of $\lvert q \rvert = 48 \pm 9 \ e_0$ (charge to mass ratio $\lvert q/m \rvert = 2.8 \pm 0.6 \ \text{C}/\text{kg}$). For a $1$~kV applied voltage we find $F=2.7 \pm 0.5 \times 10^{-15}$~N. In principle, much smaller forces could be detected, which warrants further experimental investigation. For more sensitive characterization of static perturbations, further data at $<$ 1~kV voltage on the needle is needed (or a smaller particle charge). 

The needle setup can also be used for electrical state-based feedback control, such as cooling, of the motion of a charged nanoparticle~\cite{Goldwater2018, Iwasaki2018}. A single needle can be used to drive or cool the motion of the three translational degrees of freedom. However, by adding a further two needles each degree of freedom could be controlled independently by using the combination of the three fields to have our electric field vector pointing in any desired direction. Currently, we implement parametric feedback control by modulating power of the trapping laser. The same laser is used for detection, meaning the feedback signal is encoded in the detected signal. In principle, the needle setup is advantageous as it allows us to implement linear feedback at the oscillator frequency without mixing such a signal with the particle's detected motion. Similar feedback schemes have been used in cavity optomechanics and cantilever systems, where modulation of radiation pressure \cite{Poggio2007, Kleckner2006, Schliesser2006} and ac magnetic fields \cite{Kim2017} have been used to cool the oscillator mode.

In conclusion, we have experimentally demonstrated Fano anti-resonance in levitated optomechanics by introducing an electrostatic perturbation to a charged levitated nanoparticle's potential, with the ability to tune the Fano parameter by varying the applied voltage on a nearby charged needle. We have experimentally extracted the charge to mass ratio of trapped nanoparticles. Furthermore, we have shown that we can use the induced Fano parameter as a tool to quantify static interactions which perturb the nanoparticle's trapping potential, with a static detectable force of $2.7 \pm 0.5 \times 10^{-15}$~N reported. Although our results aren't yet comparable to the 10~aN static force sensitivity reported elsewhere~\cite{Hebestreit2018}, we note that our method can measure the force in one second, compared to averaging the results of thousands of free fall experiments. As a sensor, this could be used as a tool to measure short range interactions, such as the Casimir force, or for sensitive gravity detection. The source of $\gamma_{\text{el}}$ is currently unknown, but it could be due to vibrational phonons inside the silica nanoparticle \cite{Heiman1979}.

We would like to acknowledge A. Setter, A. Vinante and L. Ferialdi for discussions, and G. Savage and P. Connell for their technical expertise. We would also like to thank the Leverhulme Trust {[}RPG-2016-046{]} and the EU Horizon 2020 research and innovation programme under grant agreement No 766900 {[}TEQ{]} for funding support. All data supporting this study are openly available from the University of Southampton repository at https://doi.org/10.5258/SOTON/D0713.

\bibliographystyle{apsrev4-1}
\bibliography{mendeley.bib}

\end{document}